\documentclass[10pt]{article}
\usepackage{multicol}
\usepackage{graphicx}
\usepackage{amsmath}
\usepackage[a4paper]{geometry}
\usepackage{hyperref}
\usepackage{rotating}

\begin{document}

\begin{center}
{\Large Energy scan/dependence of kinetic freeze-out scenarios of
multi-strange and other identified particles in central
nucleus-nucleus collisions}

\vskip1.0cm

Muhammad~Waqas$^{1,2,3,}${\footnote{E-mail:
waqas\_phy313@yahoo.com}},
Fu-Hu~Liu$^{1,2,}${\footnote{Corresponding author. E-mail:
fuhuliu@163.com; fuhuliu@sxu.edu.cn }},
Rui-Qin~Wang$^{4,}${\footnote{E-mail: wangrq@qfnu.edu.cn}},
Irfan~Siddique$^{5,}${\footnote{E-mail: irfans@mail.ustc.edu.cn}}
\\

{\small\it $^1$Institute of Theoretical Physics \& State Key
Laboratory of Quantum Optics and Quantum Optics Devices,\\ Shanxi
University, Taiyuan, Shanxi 030006, People's Republic of China

$^2$Collaborative Innovation Center of Extreme Optics, Shanxi
University,\\ Taiyuan, Shanxi 030006, People's Republic of China

$^3$School of Nuclear Science and Technology, University of
Chinese Academy of Sciences,\\ Beijing 100049, People's Republic
of China

$^4$Department of Physics, Qufu Normal University, Qufu, Shandong
273165, People's Republic of China

$^5$Department of Modern Physics, University of Science and
Technology of China,\\ Hefei, Anhui 230026, People's Republic of
China}

\end{center}

\vskip1.0cm

{\bf Abstract:} The transverse momentum (mass) spectra of the
multi-strange and non-multi-strange (i.e. other identified)
particles in central gold-gold (Au-Au), lead-lead (Pb-Pb),
argon-muriate (Ar-KCl) and nickel-nickel (Ni-Ni) collisions over a
wide energy range have been studied in this work. The experimental
data measured by various collaborations have been analyzed. The
blast-wave fit with Tsallis statistics is used to extract the
kinetic freeze-out temperature and transverse flow velocity from
the experimental data of transverse momentum (mass) spectra. The
extracted parameters increase with the increase of collision
energy and appear with the trend of saturation at the Beam Energy
Scan (BES) energies at the Relativistic Heavy Ion Collider (RHIC).
This saturation implies that the onset energy of phase transition
of partial deconfinement is 7.7 GeV and that of whole
deconfinement is 39 GeV. Furthermore, the energy scan/dependence
of kinetic freeze-out scenarios are observed for the multi-strange
and other identified particles, though the multiple freeze-out
scenarios are also observed for various particles.
\\



\vskip1.0cm
\begin{multicols}{2}

{\section{Introduction}}

High energy heavy ion collisions at the Relativistic Heavy Ion
Collider (RHIC) in the Beam Energy Scan (BES) program offer a
unique possibility to explore the quantum chromodynamics (QCD)
phase diagram~\cite{1,2,3,4}. The usual phase diagram of QCD is
plotted as the chemical freeze-out temperature ($T_{ch}$) versus
baryon chemical potential $\mu_B$. Let us assume a thermalized
system which is created in heavy ion collisions, $T_{ch}$ and
$\mu_B$ are expected to be varied with changing the collision
energy~\cite{5,6,7}. Theories suggest the formation of QCD phase
diagram which includes a possible transition from a high density
and high temperature phase known as quark-gluon plasma (QGP) phase
and this phase has been predicted by the lattice QCD~\cite{8}.

Lattice QCD calculations indicate the evolution of a rapid
cross-over at the hadron to parton phase transition~\cite{9,10} in
the system at $\mu_B=0$. Several QCD-based
models~\cite{11,12,13,13a} and the calculations from lattice
QCD~\cite{11} suggest the first order phase transition if the
system created in collisions correspond to larger values of
$\mu_B$. The point in QCD phase diagram plane where the first
order phase transition ends, is known as the QCD critical
point~\cite{14,15}.

Experimental and theoretical nuclear physics research is currently
focused on the digging out of critical point and the phase
boundary in the QCD phase diagram. The RHIC has undertaken the
first phase of the BES program~\cite{16,17,18,19,20} upto this
end, by varying the collision energy from the top RHIC to the
lower most possible energy in order to look for the signatures of
QCD phase boundary.

However, before looking to these signatures, it is very important
to know the $T_{ch}$--$\mu_B$ region of the phase diagram we can
access at the chemical freeze-out, as well as the kinetic
freeze-out temperature $T_0$ or $T_{kin}$ and transverse flow
velocity $\beta_T$ at the thermal and kinetic freeze-out. The
transverse momentum ($p_T$) or mass ($m_T$) distributions of the
particles and their yields are the tool to study the thermal and
collective properties of the dense and hot hadronic matter formed
in high energy collisions and it allows us to infer the related
parameters at the kinetic and chemical freeze-outs.

In general, the freeze-out itself may be a complicated process, as
it involves the duration in time, and a hierarchy where different
types of particles and different reactions switch off at different
times. According to the general kinematic arguments, it is
expected that the reactions with the lower collision frequency
then lower total cross-section switches off at higher
densities/temperatures. However, the one with larger collision
frequency and then larger total cross-section lasts longer.
Therefore, the elastic cross-section is larger than inelastic
cross-section in most cases, so the earlier occurrence of the
inelastic (chemical) freeze-out than the elastic (kinetic)
freeze-out is expected to happen. Generally, $T_{ch}>T_0$
according to the early study on this topic in ref.~\cite{21}.

Furthermore, it is understood that the temperature is surely one
of the most central concepts in thermodynamics and statistical
mechanics. Due to it's extremely wide applications on experimental
measurements and theoretical studies, temperature is very
important concept in both the thermal and sub-atomic physics. At
least four types of temperatures can be found in literature of
physics of high energy collisions, which includes the initial
temperature, chemical freeze-out temperature, kinetic freeze-out
temperature and effective temperature. These temperatures occur at
different stages of the collision process. We have discussed some
of these temperatures in our previous works~\cite{22,23,24,24a}.

Three types of different freeze-out scenarios can be found in
literature which includes single~\cite{25}, double~\cite{26,27}
and multiple kinetic freeze-out scenarios~\cite{28,29}. In single
freeze-out scenario, one set of parameters should be used for both
the spectra of strange and non-strange particles~\cite{25}, while
one set of parameters for strange (or multi-strange) particles and
other for non-strange (or non-multi-strange) particles should be
used in double kinetic freeze-out scenario~\cite{26,27}. Different
sets of parameters for different particles with different masses
should be used for multiple kinetic freeze-out
scenario~\cite{28,29}. It is needed to find out that which
freeze-out scenario is correct.

In this article, we are focusing on the kinetic freeze-out
temperature $T_0$ and transverse flow velocity $\beta_T$. We are
interested in the onset energy of phase transition and the kinetic
freeze-out scenario which can be obtained from the analysis of
$T_0$ and $\beta_T$. We shall extract the two parameters from the
transverse momentum (mass) spectra of different particles by using
the blast-wave fit with Tsallis statistics.

The remainder of the paper includes the method and formalism,
results and discussion, as well as summary and conclusions which
are presented in sections 2, 3 and 4, respectively.
\\

{\section{The method and formalism}}

We discuss the complex process of high energy collisions in the
framework of the blast-wave fit with Tsallis statistics~\cite{25}.
Various distributions can be used to describe the multiple
emission sources and the complex structure of $p_T$ spectra, which
include but are not limited to the Erlang distribution~\cite{30},
the standard distribution (Boltzmann, Fermi-Dirac and
Bose-Einstein distributions)~\cite{31}, the Tsallis
distribution~\cite{32,33,34}, the Tsallis+standard
distribution~\cite{35,36,37,38,39,40}, the Schwinger
mechanism~\cite{41,42}, the blast-wave fit with boltzmann
statistics~\cite{43,44,45,46} and so forth. These can be the
choices in the soft excitation process. Although the probability
density function can be of various forms, it is not enough to
describe the $p_T$ spectra, particularly the maximum $p_T$ up to
100 GeV/c in collisions at LHC~\cite{47}.

In fact, several $p_T$ regions, including the first region with
$p_T<4$--6 GeV/$c$, the second region with 4--6 GeV/$c$
$<p_T<17$--20 GeV/$c$ and the third region with $p_T>17$--20
GeV/$c$ have been observed in ref.~\cite{48}. Different $p_T$
regions are expected to correspond to different mechanisms.
According to ref.~\cite{48}, different whole features of
fragmentation and hadronization of partons through the string
dynamics can be reflected by different $p_T$ regions. The effect
and changes by the medium in the first $p_T$ region take part in
the main role, while in second $p_T$ region, it appears weakly.
The third $p_T$ region reflects the negligible influence of the
medium on the nuclear transparency. Maximum number of strings are
expected to have in the second $p_T$ region from every point of
view and it results in fusion and creation of strings and the
partons collective behavior.

We would like to point out that we mention about different $p_T$
regions corresponding to different mechanisms. This may be
translated into having different fitting functions with different
parameters when one fits the spectra in these regions. However, on
the other hand, there are universality, similarity or common
characteristic in high energy
collisions~\cite{48a,48b,48c,48d,48e,48e1,48e2,48e3}. This means
that one may also use the same function to fit the spectra in wide
$p_T$ range with the same fitting parameters. In fact, in
ref.~\cite{48f}, the Tsallis-like distribution is able to fit the
ATLAS and CMS spectra over 14 orders of magnitude with the same
value of the parameters. Indeed, there are both the particular and
common characteristics in high energy collisions.

Although various distributions may be used for describing the
particle spectra, the Tsallis distribution and its alternative
forms can fit the wider spectra. In particular, the Tsallis
distribution can cover the two- or three-component standard
distribution~\cite{48g}. This means that, to fit the spectra as
widely as possible, one may use the Tsallis distribution and its
alternative forms. In addition, the blast-wave fit can be easily
used to extract synchronously the kinetic freeze-out temperature
$T_0$ and transverse flow velocity $\beta_T$. Thus, it is
convenient for us to use the combination of the Tsallis
distribution and the blast-wave fit to extract $T_0$ and $\beta_T$
from enough wide spectra. This combination is in fact the the
blast-wave fit with Tsallis statistics~\cite{25,49}.

According to~\cite{25}, the probability density function of $p_T$
at mid-rapidity in the blast-wave fit with Tsallis statistics can
be given by
\begin{align}
f_S(p_T)=& C p_T m_T \int_{-\pi}^\pi d\phi\int_0^R rdr \bigg\{{1+\frac{q-1}{T_0}} \nonumber\\
&\times \bigg[m_T  \cosh(\rho)-p_T \sinh(\rho)
\cos(\phi)\bigg]\bigg\}^{-\frac{q}{q-1}},
\end{align}
where $C$ is the normalization constant, which results in the
integral of Eq. (1) to be normalized to 1,
$m_T=\sqrt{p_T^2+m_0^2}$ and $m_0$ is the rest mass, $\phi$ and
$r$ denote the azimuthal angle and radial coordinate respectively,
$R$ is the maximum $r$, $q$ is the entropy index, $T_0$ is the
kinetic freeze out temperature, $\rho=\tanh^{-1}[\beta(r)]$ is the
boost angle, $\beta(r)=\beta_S(r/R)^{n_0}$ is a self-similar flow
profile, $\beta_S$ is the flow velocity on the surface and $n_0=1$
according to ref.~\cite{25}. Particularly $\beta_T= (2/R^2)
\int_0^R r\beta(r)dr= 2\beta_S/(n_0+2)= 2\beta_S/3$.

It should be noted that the index $q/(q-1)$~\cite{49a} used in Eq.
(1) is a replacement of $1/(q-1)$ used in refs.~\cite{25,49} due
to the fact that $q$ is also required for the thermodynamic
consistency~\cite{34,49a}. Because of $q$ being close to 1, this
replacement causes a very small difference in the values of $q$ in
the two cases. In addition, Eq. (1) is valid only at around
mid-rapidity due to the fact that we have used $m_T\cosh y \approx
m_T$ to simply the equation from the integral of $y$. This
simplification affects the normalization constant $C$, but there
is no obvious influence on $T_0$ and $\beta_T$ due to narrow
mid-rapidity range being used. If the rapidity range is not around
$y=0$, we may transform it to around $y=0$ to subtract the
influence of longitudinal motion of emission source. As the
normalization constant for the probability density function Eq.
(1), $C$ does not affect the free parameters.

In this paper, we use the blast-wave fit with Tsallis statistics
to describe the soft excitation process. But in some cases the fit
in high $p_T$ region is not well, then we can use the
two-component fit in which the second component describes the hard
scattering process. The Hagedorn function~\cite{50,51} or inverse
power law~\cite{52,53,54} can be used for the second component,
that is
\begin{align}
f_H(p_T)=Ap_T\bigg(1+\frac{p_T}{p_0}\bigg)^{-n},
\end{align}
where $A$ is the normalization constant which results in the
integral of Eq. (2) to be normalized to 1, and $p_0$ and $n$ are
the free parameters.

For the harder part of the spectra, we suggest to use the Hagedorn
function given by Eq. (2) which is similar to the Tsallis-like
function without the blast-wave corrections~\cite{34,49a}. In
fact, there is a relation $n=q/(q-1)$ between the two functions.
It is acceptable if we use the Tsallis-like function instead of
the Hagedorn function in the case of the two temperatures (two
entropy indexes) being distinguished for with and without the
blast-wave corrections. The blast-wave corrections are not needed
for the harder part of the spectra due to its earlier production
than the softer part, when the blast-wave is not formed.

To describe the spectra in a wide $p_T$ range, we have two methods
to superpose Eqs. (1) and (2). That is
\begin{align}
f_0(p_T)=kf_S(p_T)+(1-k)f_H(p_T)
\end{align}
and
\begin{align}
f_0(p_T)=A_1 \theta(p_1-p_T) f_S(p_T) + A_2
\theta(p_T-p_1)f_H(p_T),
\end{align}
where $k$ ($1-k$) denotes the contribution fraction of the soft
excitation (hard scattering) process in the first method.
Naturally, the integral of Eq. (3) is normalized to 1. Meanwhile,
in the second method~\cite{50}, $A_1$ and $A_2$ are the
normalization constants which result in the two components to be
equal to each other at $p_T=p_1$. The function $\theta(x)=0$ if
$x<0$ and $\theta(x)=1$ if $x\geq0$. The integral of Eq. (4) is
normalized to 1, too. The contribution fraction $k$ ($1-k$) of the
soft excitation (hard scattering) process in the second method is
the integral of the first (second) component in Eq. (4).

To use Eq. (4) and to decide the value of $p_1$, we may fit the
low- and high-$p_T$ regions by the first and second components
respectively~\cite{24}. It is general that the first component
cannot fit the high-$p_T$ region and the second component cannot
fit the low-$p_T$ region. There is a cross connection between the
two components. The value of $p_T$ at the cross connection is
naturally regarded as the value of $p_1$. Because of the cross
connection is restricted by the two components, $p_1$ is not a
free parameter. Generally, the curve at $p_T=p_1$ is possibly not
too smooth.

If the spectra are not in a very wide $p_T$ range, the second
component in Eqs. (3) and (4) are not necessary in the fitting
procedure. Thus, we can use only the first component in Eqs. (3)
and (4), that is Eq. (1), to fit the spectra. Although Eqs.
(2)--(4) are not used in the fitting procedure in this work, we
present them to show a whole treatment in methodology. In the case
of analyzing the spectra in wide $p_T$ range, we may use together
Eqs. (1) and (2) due to Eq. (3) or (4).

In some cases, the spectra are in the form of $m_T$, but not
$p_T$. Then, we need to convert the $p_T$ distribution $f_S(p_T)$
to the $m_T$ distribution $f_{S'}(m_T)$ by
$f_{S'}(m_T)|dm_T|=f_S(p_T)|dp_T|$ through $p_T|dp_T|=m_T|dm_T|$
due to the invariant cross-section. In fact, Eq. (1) used in
ref.~\cite{25} appearing in the form of $f_{S'}(m_T)$. We convert
it to the form of $f_S(p_T)$ expediently. To extract $T_0$ and
$\beta_T$, we do not need the spectra in a wide $p_T$ range due to
small fraction in high $p_T$ region.
\\

{\section{Results and discussion}}

Figures 1(a)--1(d) and the continued part Figures 1(e)--1(h)
demonstrate the $p_T$ or $m_T-m_0$ spectra, $(1/2\pi
p_T)d^2N/dp_Tdy$, $d^2N/dp_Tdy$, $(1/2\pi m_T)d^2N/dm_Tdy$, or
$(1/m_T^2)d^2N/dm_Tdy$, of the non-multi-strange (i.e. other
identified) particles ($\pi^+$, $K^+$, $p$, $K^0_S$ and $\Lambda$)
and multi-strange particles [$\phi$, $\bar\Xi^+$ ($\Xi^+$, $\Xi$)
and $\bar \Omega^+$ ($\Omega^-+\Omega^+$, $\Omega$)], produced at
mid-rapidity (mid-$y$) or mid-pseudorapidity (mid-$\eta$) in
central Au-Au, Pb-Pb, Ar-KCl and Ni-Ni collisions at different
$\sqrt{s_{NN}}$, the center-of-mass energy per nucleon pair, where
$N$ denotes the number of particles. The particle types and
collision energies are marked in the panels. The symbols in panels
(a)--(c) represent the experimental data measured by the
E866~\cite{55}, E895~\cite{56,57}, E802~\cite{58,59},
STAR~\cite{60,61,62}, PHENIX~\cite{63,64} and ALICE
collaborations~\cite{65}. In panel (d) the symbols represent the
experimental data measured by the HADES~\cite{66,67},
STAR~\cite{68,69,70} and CMS Collaborations~\cite{71}. The symbols
in panels (e), (f), (g) and (h) represent the experimental data
quoted from
refs.~\cite{68,72,74},~\cite{68,75,76,77},~\cite{68,76,77}
and~\cite{63,64,71,72}, respectively. The curves are our fitted
results by using Eq. (1). The values of the free parameters
($T_0$, $\beta_T$ and $q$), the normalization constant ($N_0$),
$\chi^2$, and the number of degree of freedom (ndof) are given in
Table 1 and its continued part with together the concrete
collisions, energies, centrality, (pseudo)rapidity, particles,
spectra and scaled factors in the figure. Due to the resonance
production, the spectra in very low-$p_T$ region are not taken
care carefully in the fit process, while the fit itself is not too
good. One can see that the blast-wave fit with Tsallis statistics
fits approximately the experimental data over a wide energy range.

It should be noted that the normalization constant $N_0$ is used
to compare the fit function $f_S(p_T)$ (or $f_{S'}(m_T)$) and the
experimental spectra, and the normalization constant $C$ is used
to let the integral of Eq. (1) be 1. The two normalization
constants are different, though $C$ can be absorbed in $N_0$. We
have used both the $C$ and $N_0$ to give a clear description. In
the comparisons, we have $(1/2\pi p_T)N_0f_S(p_T)/dy=(1/2\pi
p_T)d^2N/dp_Tdy$, $N_0f_S(p_T)/dy=d^2N/dp_Tdy$, $(1/2\pi
m_T)N_0f_{S'}(m_T)/dy=(1/2\pi m_T)d^2N/dm_Tdy$, or
$(1/m_T^2)N_0f_{S'}(m_T)/dy=(1/m_T^2)d^2N/dm_Tdy$, due to
different forms of the spectra. In particular, the value of $N_0$
for $K_S^0$ production in Ar-KCl collisions at 2.25 GeV is very
small due to less participant nucleons at lower collision energy
performed in ref.~\cite{66}.

\begin{figure*}[htb!]
\begin{center}
\vskip3cm \hskip-2.8cm
\includegraphics[width=10.5cm]{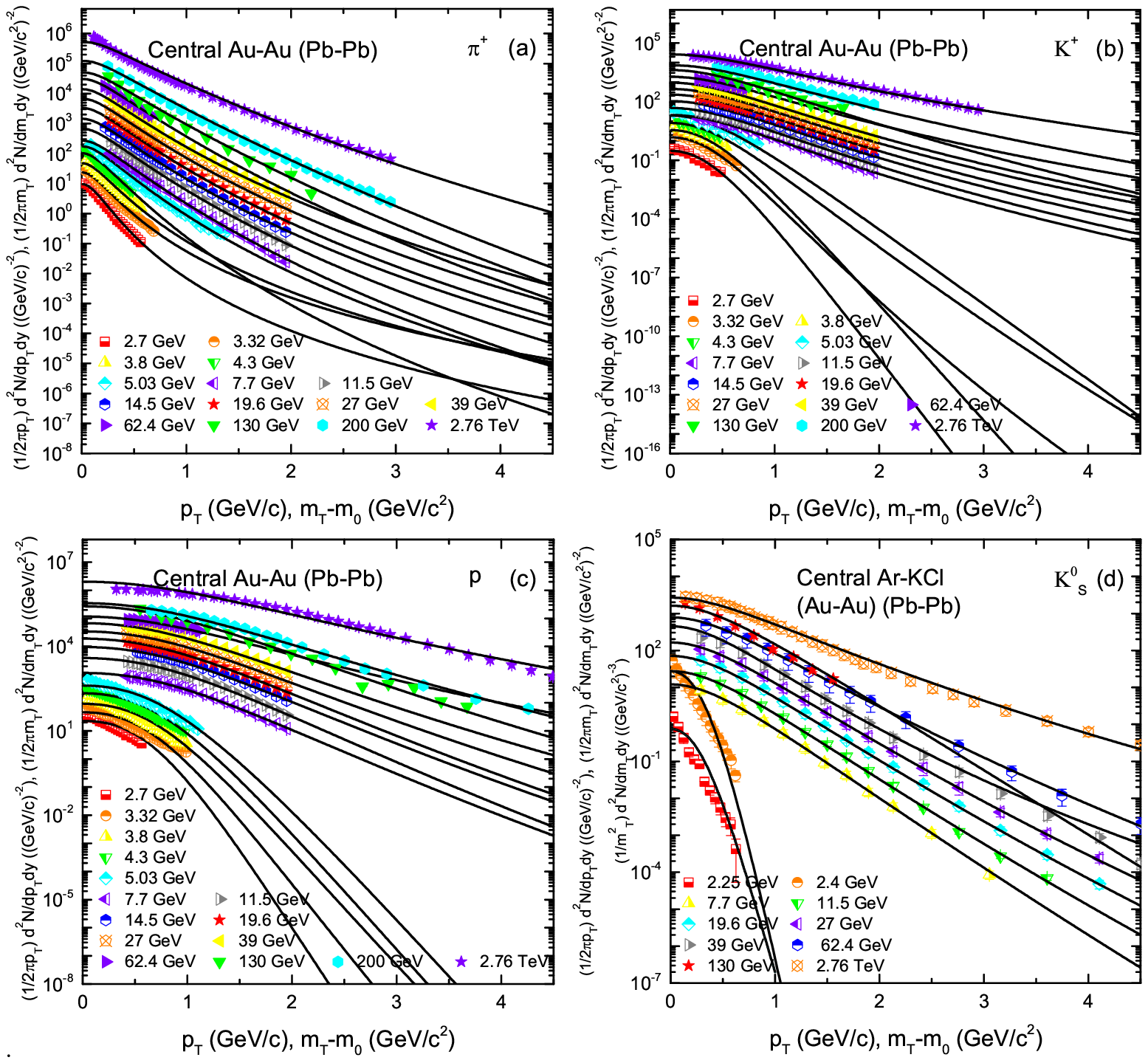}
\end{center}
\vskip-1cm {\small Fig. 1. The $p_T$ or $m_T-m_0$ spectra,
$(1/2\pi p_T)d^2N/dp_Tdy$, $(1/2\pi m_T)d^2N/dm_Tdy$,
$d^2N/dp_Tdy$, or $(1/m_T^2)d^2N/dm_Tdy$, of other identified
particles ($\pi^+$, $K^+$, $p$ and $K^0_S$), produced at mid-$y$
in central Au-Au, Pb-Pb and Ar-KCl collisions at different
$\sqrt{s_{NN}}$. The particle types and collision energies are
marked in the panels. The symbols in panels (a)--(c) represent the
experimental data measured by the E866~\cite{55},
E895~\cite{56,57}, E802~\cite{58,59}, STAR~\cite{60,61,62},
PHENIX~\cite{63,64} and ALICE collaborations~\cite{65}. In panel
(d) the symbols represent the experimental data measured by the
HADES~\cite{66,67}, STAR~\cite{68,69,70} and CMS
Collaborations~\cite{71}. The curves are our fitted results by
using Eq. (1). More information in detail can be found in Table
1.}
\end{figure*}

\begin{figure*}[htb!]
\begin{center}
\includegraphics[width=13.cm]{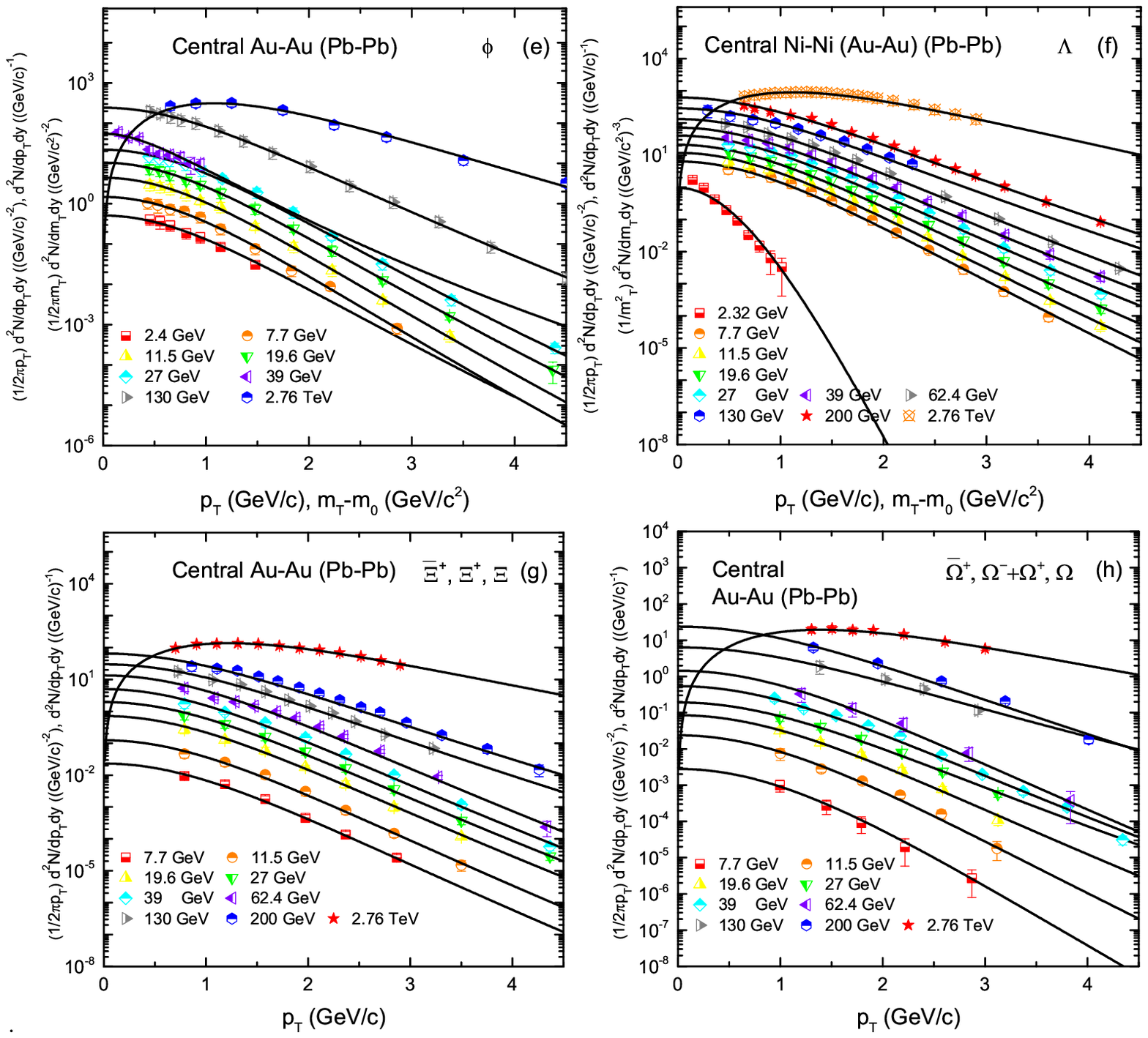}
\end{center}
{\small Fig.1. Continued. The same as for Figure 1, but showing
the $p_T$ or $m_T-m_0$ spectra of other identified particles
($\Lambda$) and multi-strange particles [$\phi$, $\bar\Xi^+$
($\Xi^+$, $\Xi$) and $\bar \Omega^+$ ($\Omega^-+\Omega^+$,
$\Omega$)] produced at mid-$y$ in central Au-Au, Pb-Pb and Ni-Ni
collisions at different $\sqrt{s_{NN}}$. The symbols in panels
(e), (f), (g) and (h) represent the experimental data quoted from
refs.~\cite{68,72,74},~\cite{68,75,76,77},~\cite{68,76,77}
and~\cite{63,64,71,72}, respectively. More information in detail
can be found in Table 1 continued part.}
\end{figure*}

\end{multicols}
\begin{sidewaystable}
{\tiny Table 1. Values of $T_0$, $\beta_T$, $q$, $N_0$, $\chi^2$,
and ndof corresponding to the curves in Fig. 1, where the
centrality classes, (pseudo)rapidity ranges, types of spectra and
the scaled factors are listed. The scaled factors are just used
for the display purpose only. \vspace{-.25cm}
\begin{center}
\begin{tabular}{ccccccccccccc}\\ \hline\hline
Figure & Energy & Centrality & $y$ ($\eta$) & Particle & Spectrum
& Scaled & $T_0$ (GeV) &$\beta_T(c)$ & $q$ & $N_0$ & $\chi^2$ &
ndof\\\hline
Fig. 1(a)  & 2.7 GeV   & 0--5\%  & $|y|<0.05$ & $\pi^+$  & $(1/2\pi m_T)d^2N/dm_Tdy$ & 1/4 & $0.024\pm0.005$ & $0.339\pm0.009$ & $1.130\pm0.020$ &  $0.80\pm0.05$    & 3   & 22\\
 Au-Au     & 3.32 GeV  & 0--5\%  & $|y|<0.05$ &          & $(1/2\pi m_T)d^2N/dm_Tdy$ & 1/3 & $0.031\pm0.004$ & $0.344\pm0.008$ & $1.150\pm0.030$ &  $2.07\pm0.20$    & 2   & 27\\
           & 3.8 GeV   & 0--5\%  & $|y|<0.05$ &          & $(1/2\pi m_T)d^2N/dm_Tdy$ & 1/2 & $0.038\pm0.006$ & $0.358\pm0.015$ & $1.120\pm0.010$ &  $3.08\pm0.20$    & 33  & 22\\
           & 4.3 GeV   & 0--5\%  & $|y|<0.05$ &          & $(1/2\pi m_T)d^2N/dm_Tdy$ & --  & $0.050\pm0.004$ & $0.376\pm0.016$ & $1.060\pm0.010$ &  $3.28\pm0.33$    & 50  & 19\\
           & 5.03 GeV  & 0--5\%  & $0<y<0.4$  &          & $(1/2\pi m_T)d^2N/dm_Tdy$ & 2   & $0.060\pm0.005$ & $0.390\pm0.010$ & $1.086\pm0.005$ &  $7.10\pm1.00$    & 96  & 34\\
           & 7.7 GeV   & 0--5\%  & $|y|<0.1$  &          & $(1/2\pi p_T)d^2N/dp_Tdy$ & --  & $0.068\pm0.004$ & $0.429\pm0.008$ & $1.058\pm0.005$ &  $15.33\pm2.00$   & 113 & 26\\
           & 11.5 GeV  & 0--5\%  & $|y|<0.1$  &          & $(1/2\pi p_T)d^2N/dp_Tdy$ & 2   & $0.067\pm0.005$ & $0.430\pm0.008$ & $1.069\pm0.006$ &  $19.43\pm2.80$   & 119 & 26\\
           & 14.5 GeV  & 0--5\%  & $|y|<0.1$  &          & $(1/2\pi p_T)d^2N/dp_Tdy$ & 4   & $0.068\pm0.004$ & $0.428\pm0.006$ & $1.080\pm0.007$ &  $21.93\pm2.20$   & 153 & 28\\
           & 19.6 GeV  & 0--5\%  & $|y|<0.1$  &          & $(1/2\pi p_T)d^2N/dp_Tdy$ & 8   & $0.068\pm0.005$ & $0.429\pm0.009$ & $1.058\pm0.028$ &  $25.20\pm4.00$   & 178 & 26\\
           & 27 GeV    & 0--5\%  & $|y|<0.1$  &          & $(1/2\pi p_T)d^2N/dp_Tdy$ & 16  & $0.069\pm0.004$ & $0.430\pm0.009$ & $1.080\pm0.006$ &  $26.43\pm2.25$   & 49  & 26\\
           & 39 GeV    & 0--5\%  & $|y|<0.1$  &          & $(1/2\pi p_T)d^2N/dp_Tdy$ & 32  & $0.069\pm0.005$ & $0.430\pm0.008$ & $1.080\pm0.007$ &  $28.43\pm3.50$   & 93  & 26\\
           & 62.4 GeV  & 0--5\%  & $|y|<0.1$  &          & $(1/2\pi p_T)d^2N/dp_Tdy$ & 64  & $0.076\pm0.006$ & $0.447\pm0.007$ & $1.040\pm0.008$ &  $32.53\pm2.50$   & 63  & 10 \\
           & 130 GeV   & 0--5\%  & $|\eta|<0.35$&        & $(1/2\pi p_T)d^2N/dp_Tdy$ & 120 & $0.082\pm0.006$ & $0.478\pm0.008$ & $1.031\pm0.006$ &  $34.43\pm2.70$   & 3   & 14\\
           & 200 GeV   & 0--5\%  & $|\eta|<0.35$&        & $(1/2\pi p_T)d^2N/dp_Tdy$ & 240 & $0.089\pm0.005$ & $0.489\pm0.009$ & $1.024\pm0.008$ &  $95.80\pm10.00$  & 94  & 28\\
 Pb-Pb     & 2.76 TeV  & 0--5\%  & $|y|<0.5$  &          & $(1/2\pi p_T)d^2N/dp_Tdy$ & --  & $0.094\pm0.006$ & $0.500\pm0.011$ & $1.040\pm0.007$ &  $122.00\pm10.00$ & 90  & 41\\
\hline
Fig. 1(b)  & 2.7 GeV   & 0--5\%  &$|y|<0.23$ & $K^+$     & $(1/2\pi m_T)d^2N/dm_Tdy$ & --  & $0.027\pm0.006$ & $0.300\pm0.015$ & $1.001\pm0.008$ &  $0.014\pm0.005$ & 1  & 10\\
 Au-Au     & 3.32 GeV  & 0--5\%  &$|y|<0.29$ &           & $(1/2\pi m_T)d^2N/dm_Tdy$ & --  & $0.033\pm0.006$ & $0.317\pm0.016$ & $1.014\pm0.005$ &  $0.080\pm0.004$ & 1  & 12\\
           & 3.8 GeV   & 0--5\%  &$|y|<0.34$ &           & $(1/2\pi m_T)d^2N/dm_Tdy$ & 1.2 & $0.040\pm0.007$ & $0.331\pm0.012$ & $1.001\pm0.008$ &  $0.20\pm0.04$   & 5  & 11\\
           & 4.3 GeV   & 0--5\%  &$|y|<0.37$ &           & $(1/2\pi m_T)d^2N/dm_Tdy$ & 2   & $0.048\pm0.006$ & $0.347\pm0.012$ & $1.001\pm0.008$ &  $0.30\pm0.05$   & 4  & 9\\
           & 5.03 GeV  & 0--5\%  &$|y|<0.1$  &           & $(1/2\pi m_T)d^2N/dm_Tdy$ & 3   & $0.059\pm0.005$ & $0.362\pm0.010$ & $1.001\pm0.007$ &  $0.60\pm0.03$   & 11 & 11\\
           & 7.7 GeV   & 0--5\%  &$|y|<0.1$  &           & $(1/2\pi p_T)d^2N/dp_Tdy$ & --  & $0.072\pm0.004$ & $0.397\pm0.007$ & $1.055\pm0.006$ &  $3.28\pm0.40$   & 9  & 23\\
           & 11.5 GeV  & 0--5\%  &$|y|<0.1$  &           & $(1/2\pi p_T)d^2N/dp_Tdy$ & 2   & $0.073\pm0.005$ & $0.398\pm0.004$ & $1.055\pm0.007$ &  $4.00\pm0.50$   & 32  & 25\\
           & 14.5 GeV  & 0--5\%  &$|y|<0.1$  &           & $(1/2\pi p_T)d^2N/dp_Tdy$ & 4   & $0.073\pm0.005$ & $0.398\pm0.004$ & $1.061\pm0.007$ &  $4.33\pm0.50$   & 67  & 26\\
           & 19.6 GeV  & 0--5\%  &$|y|<0.1$  &           & $(1/2\pi p_T)d^2N/dp_Tdy$ & 8   & $0.073\pm0.006$ & $0.397\pm0.008$ & $1.065\pm0.008$ &  $4.73\pm0.70$   & 11  & 26\\
           & 27 GeV    & 0--5\%  &$|y|<0.1$  &           & $(1/2\pi p_T)d^2N/dp_Tdy$ & 16  & $0.073\pm0.006$ & $0.398\pm0.007$ & $1.066\pm0.009$ &  $4.80\pm0.55$   & 19  & 26\\
           & 39 GeV    & 0--5\%  &$|y|<0.1$  &           & $(1/2\pi p_T)d^2N/dp_Tdy$ & 32  & $0.073\pm0.004$ & $0.398\pm0.007$ & $1.056\pm0.007$ &  $5.00\pm0.60$   & 12 & 26\\
           & 62.4 GeV  & 0--5\%  &$|y|<0.1$  &           & $(1/2\pi p_T)d^2N/dp_Tdy$ & 64  & $0.079\pm0.005$ & $0.415\pm0.008$ & $1.064\pm0.385$ &  $5.93\pm0.50$   & 1  & 10\\
           & 130 GeV   & 0--5\%  &$|\eta|<0.35$&         & $(1/2\pi p_T)d^2N/dp_Tdy$ & 120 & $0.084\pm0.005$ & $0.423\pm0.011$ & $1.035\pm0.010$ &  $6.33\pm0.70$   & 12 & 13\\
           & 200 GeV   & 0--5\%  &$|\eta|<0.35$&         & $(1/2\pi p_T)d^2N/dp_Tdy$ & 340 & $0.090\pm0.005$ & $0.440\pm0.009$ & $1.052\pm0.100$ &  $6.60\pm0.60$   & 1  & 16\\
  Pb-Pb    & 2.76 TeV  & 0--5\%  &$|y|<0.5$  &           & $(1/2\pi p_T)d^2N/dp_Tdy$ & --  & $0.096\pm0.005$ & $0.460\pm0.009$ & $1.090\pm0.100$ &  $15.60\pm0.60$  & 23 & 36\\
\hline
Fig. 1(c)  & 2.7 GeV   & 0--5\%  &$|y|<0.05$ & $p$ & $(1/2\pi m_T)d^2N/dm_Tdy$ & --   & $0.030\pm0.005$ & $0.280\pm0.008$ & $1.010\pm0.006$ &  $2.53\pm0.16$  & 72  & 39\\
Au-Au      & 3.32 GeV  & 0--5\%  &$|y|<0.05$ &     & $(1/2\pi m_T)d^2N/dm_Tdy$ & 2.5  & $0.035\pm0.004$ & $0.291\pm0.007$ & $1.010\pm0.010$ &  $2.34\pm0.16$  & 57  & 39\\
           & 3.8 GeV   & 0--5\%  &$|y|<0.05$ &     & $(1/2\pi m_T)d^2N/dm_Tdy$ & 6    & $0.041\pm0.008$ & $0.307\pm0.016$ & $1.010\pm0.006$ &  $2.27\pm0.16$  & 151 & 39\\
           & 4.3 GeV   & 0--5\%  &$|y|<0.05$ &     & $(1/2\pi m_T)d^2N/dm_Tdy$ & 16   & $0.062\pm0.006$ & $0.317\pm0.013$ & $1.005\pm0.006$ &  $2.16\pm0.10$  & 73  & 36\\
           & 5.03 GeV  & 0--5\%  &$0<y<0.2$  &     & $(1/2\pi m_T)d^2N/dm_Tdy$ & 30   & $0.057\pm0.007$ & $0.331\pm0.011$ & $1.001\pm0.007$ &  $2.43\pm0.20$  & 49  & 29\\
           & 7.7 GeV   & 0--5\%  &$|y|<0.1$  &     & $(1/2\pi p_T)d^2N/dp_Tdy$ & 50   & $0.075\pm0.007$ & $0.407\pm0.010$ & $1.030\pm0.004$ &  $8.59\pm0.60$  & 12   & 29\\
           & 11.5 GeV  & 0--5\%  &$|y|<0.1$  &     & $(1/2\pi p_T)d^2N/dp_Tdy$ & 200  & $0.075\pm0.005$ & $0.407\pm0.007$ & $1.027\pm0.010$ &  $7.13\pm0.40$  & 8   & 28\\
           & 14.5 GeV  & 0--5\%  &$|y|<0.1$  &     & $(1/2\pi p_T)d^2N/dp_Tdy$ & 600  & $0.075\pm0.005$ & $0.408\pm0.008$ & $1.034\pm0.006$ &  $6.18\pm0.27$  & 10   & 25\\
           & 19.6 GeV  & 0--5\%  &$|y|<0.1$  &     & $(1/2\pi p_T)d^2N/dp_Tdy$ & 1300 & $0.076\pm0.004$ & $0.407\pm0.007$ & $1.033\pm0.010$ &  $5.61\pm0.50$  & 21   & 29\\
           & 27 GeV    & 0--5\%  &$|y|<0.1$  &     & $(1/2\pi p_T)d^2N/dp_Tdy$ & 2900 & $0.076\pm0.006$ & $0.407\pm0.008$ & $1.041\pm0.010$ &  $4.83\pm0.60$  & 11   & 23\\
           & 39 GeV    & 0--5\%  &$|y|<0.1$  &     & $(1/2\pi p_T)d^2N/dp_Tdy$ & 7000 & $0.076\pm0.006$ & $0.407\pm0.008$ & $1.048\pm0.010$ &  $4.13\pm0.60$  & 16   & 22\\
           & 62.4 GeV  & 0--5\%  &$|y|<0.1$  &     & $(1/2\pi p_T)d^2N/dp_Tdy$ & 14000& $0.081\pm0.004$ & $0.411\pm0.008$ & $1.080\pm0.012$ &  $4.68\pm0.27$  & 3   & 15\\
           & 130 GeV   & 0--5\%  &$|\eta|<0.35$&   & $(1/2\pi p_T)d^2N/dp_Tdy$ & 28000& $0.086\pm0.006$ & $0.418\pm0.009$ & $1.040\pm0.007$ &  $4.33\pm0.30$  & 12  & 17\\
           & 200 GeV   & 0--5\%  &$|\eta|<0.35$&   & $(1/2\pi p_T)d^2N/dp_Tdy$ & 1/20 & $0.091\pm0.004$ & $0.428\pm0.007$ & $1.050\pm0.008$ &  $4.33\pm0.33$  & 10  & 22\\
 Pb-Pb     & 2.76 TeV  & 0--5\%  &$|y|<0.5$  &     & $(1/2\pi p_T)d^2N/dp_Tdy$ & 500  & $0.098\pm0.004$ & $0.441\pm0.007$ & $1.025\pm0.008$ &  $5.02\pm0.33$  & 4  & 37\\
\hline
Fig. 1(d) Ar-KCl& 2.25 GeV  & 0--35\% & $|y|<0.05$ & $K^0_S$&$(1/ m^2_T)d^2N/dydm_T$  &$10^7$&$0.019\pm0.007$&$0.232\pm0.012$ &$1.010\pm0.007$&$0.0000090\pm0.0000010$ & 33 & 13\\
       Au-Au   & 2.4 GeV    & 0--40\% & $|y|<0.05$ &        &$(1/ m^2_T)d^2N/dydm_T$  &$10^6$&$0.023\pm0.006$&$0.242\pm0.010$ &$1.001\pm0.020$&$0.0060\pm0.0001$   & 41 & 16\\
               & 7.7 GeV    & 0--5\%  & $|y|<0.5$  &        &$(1/2\pi p_T)d^2N/dp_Tdy$& --   & $0.103\pm0.006$ & $0.390\pm0.015$ &$1.010\pm0.008$ &  $1.97\pm0.20$ & 1  & 12\\
               & 11.5 GeV   & 0--5\%  & $|y|<0.5$  &        &$(1/2\pi p_T)d^2N/dp_Tdy$& 2    & $0.104\pm0.005$ & $0.393\pm0.012$ &$1.015\pm0.008$ &  $2.37\pm0.10$ & 12 & 14\\
               & 19.6 GeV   & 0--5\%  & $|y|<0.5$  &        &$(1/2\pi p_T)d^2N/dp_Tdy$& 4    & $0.104\pm0.006$ & $0.393\pm0.010$ &$1.020\pm0.008$ &  $3.17\pm0.10$ & 6  & 15\\
               & 27 GeV     & 0--5\%  & $|y|<0.5$  &        &$(1/2\pi p_T)d^2N/dp_Tdy$& 8    & $0.104\pm0.007$ & $0.392\pm0.014$ &$1.025\pm0.009$ &  $3.77\pm0.10$ & 8  & 16\\
               & 39 GeV     & 0--5\%  & $|y|<0.5$  &        &$(1/2\pi p_T)d^2N/dp_Tdy$& 16   & $0.104\pm0.006$ & $0.392\pm0.014$ &$1.030\pm0.007$ &  $3.90\pm0.50$ & 6  & 16\\
               & 62.4 GeV   & 0--5\%  & $|y|<0.5$  &        &$(1/2\pi m_T)d^2N/dm_Tdy$& 132  & $0.114\pm0.007$ & $0.401\pm0.012$ &$1.030\pm0.009$ &  $5.30\pm0.40$ & 2  & 14\\
               & 130 GeV    & 0--6\%  & $|y|<0.5$  &        &$(1/2\pi m_T)d^2N/dm_Tdy$& 160  & $0.120\pm0.008$ & $0.412\pm0.013$ &$1.001\pm0.009$ &  $5.30\pm0.40$ & 26 & 9 \\
       Pb-Pb   & 2.76 TeV   & 0--5\%  & $|y|<0.5$  &        &$(1/2\pi p_T)d^2N/dp_Tdy$& 900  & $0.133\pm0.005$ & $0.432\pm0.014$ &$1.050\pm0.011$ &  $0.98\pm0.40$ & 10 & 32\\
\hline \hline
\end{tabular}%
\end{center}}
\end{sidewaystable}
\begin{multicols}{2}

\end{multicols}
\begin{sidewaystable}
{\scriptsize Table 1. Continued. Values of $T_0$, $\beta_T$, $q$,
$N_0$, $\chi^2$, and ndof corresponding to the curves in Fig. 1
continued part, where the centrality classes, (pseudo)rapidity
ranges, types of spectra and the scaled factors are listed. The
scaled factors are just used for the display purpose only.
\vspace{-.50cm}
\begin{center}
\begin{tabular}{ccccccccccccc}\\ \hline\hline
Figure & Energy & Centrality & $y$ ($\eta$) & Particle & Spectrum
& Scaled & $T_0$ (GeV) &$\beta_T(c)$ & $q$ & $N_0$ & $\chi^2$ &
ndof\\\hline
Fig. 1(e)    & 7.7 GeV    & 0--5\% & $|y|<0.5$ & $\phi$    & $(1/2\pi p_T)d^2N/dp_Tdy$ & -- & $0.108\pm0.007$ & $0.350\pm0.014$ &$1.030\pm0.010$ &  $0.20\pm0.03$       & 1   & 7\\
 Au-Au       & 11.5 GeV   & 0--5\% & $|y|<0.5$ &           & $(1/2\pi p_T)d^2N/dp_Tdy$ & 2  & $0.108\pm0.006$ & $0.351\pm0.013$ &$1.021\pm0.007$ &  $0.27\pm0.01$       & 2   & 10\\
             & 19.6 GeV   & 0--5\% & $|y|<0.5$ &           & $(1/2\pi p_T)d^2N/dp_Tdy$ & 4  & $0.109\pm0.007$ & $0.350\pm0.014$ &$1.022\pm0.007$ &  $0.41\pm0.01$       & 1   & 11\\
             & 27 GeV     & 0--5\% & $|y|<0.5$ &           & $(1/2\pi p_T)d^2N/dp_Tdy$ & 8  & $0.109\pm0.006$ & $0.351\pm0.015$ &$1.026\pm0.011$ &  $0.50\pm0.03$       & 2   & 12\\
             & 39 GeV     & 0--5\% & $|y|<0.5$ &           & $(1/2\pi p_T)d^2N/dp_Tdy$ & 16 & $0.108\pm0.008$ & $0.351\pm0.016$ &$1.030\pm0.009$ &  $0.54\pm0.01$       & 4   & 12\\
             & 130 GeV    & 0--11\%& $|y|<0.5$ &           & $(1/2\pi m_T)d^2N/dm_Tdy$ & 50 & $0.124\pm0.005$ & $0.370\pm0.013$ &$1.050\pm0.009$ &  $0.27\pm0.04$       & 14  & 9 \\
             & 200 GeV    & 0--10\%& $|y|<0.5$ &           & $(1/2\pi p_T)d^2N/dp_Tdy$ & 100& $0.133\pm0.006$ & $0.380\pm0.014$ &$1.030\pm0.008$ &  $1.24\pm0.40$       & 2   & 14 \\
 Pb-Pb       & 2.76 TeV   & 0--5\% & $|y|<0.5$ &           & $d^2N/dp_Tdy$             & 10 & $0.140\pm0.005$ & $0.391\pm0.012$ &$1.035\pm0.010$ &  $48.00\pm0.40$      & 4   & 8\\
\hline
Fig. 1(f) Ni-Ni& 2.32 GeV   &0--5\% & $-1<y<0$  &$\Lambda$ & $(1/ m^2_T)d^2N/dydm_T$   & $10^8$& $0.030\pm0.005$ & $0.199\pm0.015$ &$1.001\pm0.007$ &  $0.00076\pm0.00003$& 17  & 9\\
        Au-Au  & 7.7 GeV    &0--5\% & $|y|<0.5$ &          & $(1/2\pi p_T)d^2N/dp_Tdy$ &   --  & $0.114\pm0.006$ & $0.312\pm0.012$ &$1.015\pm0.008$ &  $2.20\pm0.20$      & 5   & 13\\
               & 11.5 GeV   &0--5\% & $|y|<0.5$ &          & $(1/2\pi p_T)d^2N/dp_Tdy$ &   2   & $0.115\pm0.006$ & $0.313\pm0.012$ &$1.018\pm0.008$ &  $2.10\pm0.10$      & 5   & 14\\
               & 19.6 GeV   &0--5\% & $|y|<0.5$ &          & $(1/2\pi p_T)d^2N/dp_Tdy$ &   4   & $0.115\pm0.005$ & $0.313\pm0.013$ &$1.022\pm0.007$ &  $1.95\pm0.10$      & 11  & 15\\
               & 27 GeV     &0--5\% & $|y|<0.5$ &          & $(1/2\pi p_T)d^2N/dp_Tdy$ &   8   & $0.114\pm0.006$ & $0.312\pm0.013$ &$1.027\pm0.009$ &  $1.95\pm0.10$      & 9   & 15\\
               & 39 GeV     &0--5\% & $|y|<0.5$ &          & $(1/2\pi p_T)d^2N/dp_Tdy$ &   16  & $0.115\pm0.007$ & $0.313\pm0.012$ &$1.035\pm0.008$ &  $1.78\pm0.50$      & 7   & 14\\
               & 62.4 GeV   &0--5\% & $|y|<0.05$&          & $(1/2\pi p_T)d^2N/dp_Tdy$ &   32  & $0.125\pm0.005$ & $0.325\pm0.012$ &$1.029\pm0.009$ &  $1.78\pm0.40$      & 9   & 12\\
               & 130 GeV    &0--5\% & $|y|<0.05$&          & $(1/2\pi p_T)d^2N/dp_Tdy$ &   64  & $0.132\pm0.006$ & $0.337\pm0.014$ &$1.040\pm0.020$ &  $2.17\pm0.40$      & 9  & 10 \\
               & 200 GeV    &0--5\% & $|y|<0.05$&          & $(1/2\pi p_T)d^2N/dp_Tdy$ &  120  & $0.140\pm0.006$ & $0.348\pm0.011$ &$1.028\pm0.030$ &  $2.67\pm0.40$      & 22  & 17 \\
       Pb-Pb   & 2.76 TeV   &0--5\% & $|y|<0.05$&          & $d^2N/dp_Tdy$             &   70  & $0.149\pm0.007$ & $0.359\pm0.012$ &$1.045\pm0.013$ &  $19.78\pm3.00$     & 27  & 19\\
\hline
Fig. 1(g)      & 7.7 GeV    &0--5\% &$|y|<0.5$ & $\bar \Xi^+$ & $(1/2\pi p_T)d^2N/dp_Tdy$  & -- & $0.119\pm0.006$ & $0.289\pm0.011$ &$1.022\pm0.011$ &  $0.010\pm0.003$     & 2   & 6\\
 Au-Au         & 11.5 GeV   &0--5\% &$|y|<0.5$ &              & $(1/2\pi p_T)d^2N/dp_Tdy$  & 2  & $0.120\pm0.005$ & $0.290\pm0.012$ &$1.022\pm0.008$ &  $0.27\pm0.01$       & 3   & 7\\
               & 19.6 GeV   &0--5\% &$|y|<0.5$ &              & $(1/2\pi p_T)d^2N/dp_Tdy$  & 4  & $0.120\pm0.005$ & $0.290\pm0.012$ &$1.025\pm0.007$ &  $0.080\pm0.001$     & 4   & 7\\
               & 27 GeV     &0--5\% &$|y|<0.5$ &              & $(1/2\pi p_T)d^2N/dp_Tdy$  & 8  & $0.120\pm0.006$ & $0.290\pm0.014$ &$1.026\pm0.012$ &  $0.11\pm0.03$       & 5   & 8\\
               & 39 GeV     &0--5\% &$|y|<0.5$ &              & $(1/2\pi p_T)d^2N/dp_Tdy$  & 16 & $0.120\pm0.006$ & $0.290\pm0.013$ &$1.026\pm0.009$ &  $0.20\pm0.008$      & 8   & 8\\
               & 62.4 GeV   &0--5\% &$|y|<0.05$& $\Xi^+$      & $(1/2\pi p_T)d^2N/dp_Tdy$  & 32 & $0.127\pm0.007$ & $0.304\pm0.013$ &$1.027\pm0.009$ &  $0.20\pm0.04$       & 31  & 11 \\
               & 130 GeV    &0--5\% &$|y|<0.05$&              & $(1/2\pi p_T)d^2N/dp_Tdy$  & 64 & $0.135\pm0.005$ & $0.315\pm0.014$ &$1.037\pm0.008$ &  $0.27\pm0.04$       & 11   & 10 \\
               & 200 GeV    &0--5\% &$|y|<0.05$&              & $(1/2\pi p_T)d^2N/dp_Tdy$  & 150& $0.142\pm0.005$ & $0.325\pm0.012$ &$1.040\pm0.008$ &  $0.27\pm0.05$       & 28  & 15 \\
  Pb-Pb        & 2.76 TeV   &0--5\% &$|y|<0.05$& $\Xi$        & $d^2N/dp_Tdy$              & 40 & $0.150\pm0.006$ & $0.334\pm0.015$ &$1.053\pm0.012$ &  $5.27\pm0.80$       & 5   & 12\\
\hline
Fig. 1(h)      & 7.7 GeV    &0--5\% &$|y|<0.5$ & $\bar \Omega^+$    & $(1/2\pi p_T)d^2N/dp_Tdy$  & -- & $0.125\pm0.005$ & $0.243\pm0.012$ &$1.012\pm0.011$ &  $0.0013\pm0.0003$   & 1   & 5\\
 Au-Au         & 11.5 GeV   &0--5\% &$|y|<0.5$ &                    & $(1/2\pi p_T)d^2N/dp_Tdy$  & 2  & $0.125\pm0.005$ & $0.243\pm0.012$ &$1.012\pm0.009$ &  $0.0060\pm0.0005$   & 6   & 6\\
               & 19.6 GeV   &0--5\% &$|y|<0.5$ &                    & $(1/2\pi p_T)d^2N/dp_Tdy$  & 4  & $0.125\pm0.005$ & $0.244\pm0.013$ &$1.030\pm0.007$ &  $0.012\pm0.002$     & 5   & 6\\
               & 27 GeV     &0--5\% &$|y|<0.5$ &                    & $(1/2\pi p_T)d^2N/dp_Tdy$  & 8  & $0.126\pm0.006$ & $0.244\pm0.014$ &$1.040\pm0.014$ &  $0.015\pm0.003$     & 4   & 6\\
               & 39 GeV     &0--5\% &$|y|<0.5$ &                    & $(1/2\pi p_T)d^2N/dp_Tdy$  & 16 & $0.126\pm0.007$ & $0.243\pm0.015$ &$1.035\pm0.008$ &  $0.020\pm0.006$     & 14  & 10\\
               & 62.4 GeV   &0--5\% &$|y|<0.05$&                    & $(1/2\pi p_T)d^2N/dp_Tdy$  & 32 & $0.134\pm0.006$ & $0.253\pm0.010$ &$1.025\pm0.008$ &  $0.26\pm0.04$       & 1   & 5 \\
               & 130 GeV    &0--5\% &$|y|<0.05$& $\Omega^-+\Omega^+$& $(1/2\pi p_T)d^2N/dp_Tdy$  & 64 & $0.142\pm0.006$ & $0.265\pm0.015$ &$1.065\pm0.008$ &  $0.090\pm0.004$     & 2   & 4 \\
               & 200 GeV    &0--5\% &$|y|<0.05$&                    & $(1/2\pi p_T)d^2N/dp_Tdy$  & 200& $0.149\pm0.007$ & $0.271\pm0.014$ &$1.045\pm0.008$ &  $2.00\pm0.20$       & 14  & 5 \\
  Pb-Pb        & 2.76 TeV   &0--5\% &$|y|<0.05$& $\Omega$           & $d^2N/dp_Tdy$              & 40 & $0.155\pm0.006$ & $0.278\pm0.014$ &$1.065\pm0.013$ &  $0.75\pm0.06$       & 2   & 7\\
\hline
\end{tabular}%
\end{center}}
\end{sidewaystable}
\begin{multicols}{2}

\begin{figure*}[htb!]
\begin{center}
\includegraphics[width=13.cm]{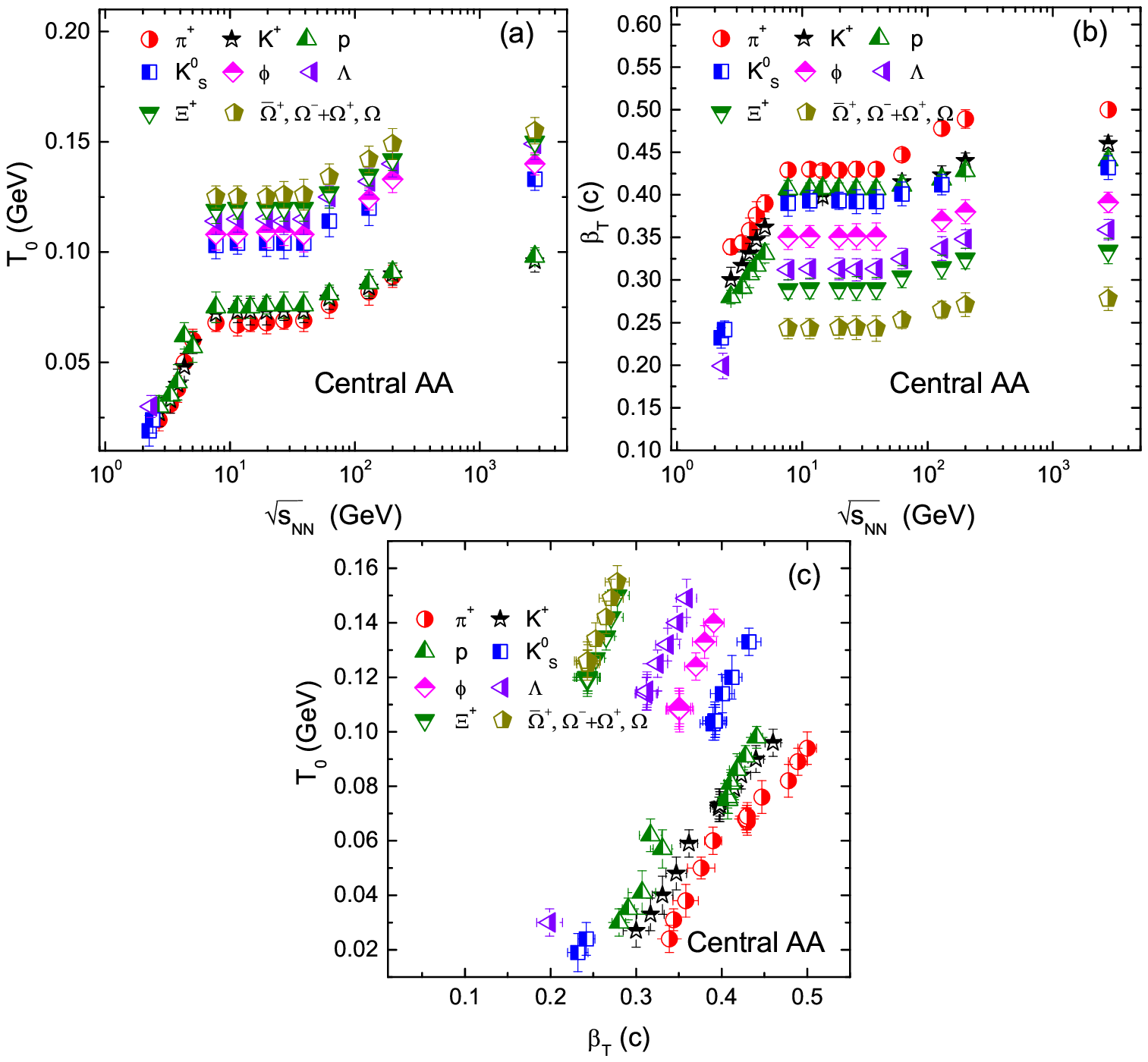}
\end{center}
Fig. 2. Dependence of (a) $T_0$ on $\sqrt{s_{NN}}$, (b) $\beta_T$
on $\sqrt{s_{NN}}$ and (c) $T_0$ on $\beta_T$ for different
$\sqrt{s_{NN}}$ in central AA collisions. The symbols marked in
the panels represent the parameter values listed in Table 1 and
its continued part.
\end{figure*}

In addition, although $R$ can be regarded as the transverse radius
of the participant region, it has no absolute meaning due to the
fact that it appears in terms of $r/R$. Although the fit result is
not related to $R$, $R$ cannot be absorbed in $C$ or $N_0$ due to
a concrete $R$ being needed to perform the calculation process. As
the simple normalization constant in probability density function
and the irrelevant upper limit in integral process, the values of
$C$ and $R$ are not listed in Table 1 to avoid trivial
presentation.

Before continuing this work, we would like to point out that Fig.
1 is only a part collection of transverse spectra. In fact, more
experimental data were published in the community. For example,
the NA49 experiment was performed for carbon-carbon (C-C),
silicon-silicon (Si-Si) and Pb-Pb collisions in which the
extensive experimental studies of the possible phase transition
were carried in the range of $\sqrt{s_{NN}}=6.3$--17.3
GeV~\cite{77a,77b,77c,77d}. The NA61/SHINE experiment was
performed at similar energies for gathering rich data on nuclear
collisions in a two-dimensional scan, i.e. varying collision
energy and nuclear size~\cite{77c,77d,77e,77f}. Indeed, these
experiments have provided more abundant data.

To study the dependence of kinetic freeze-out temperature $T_0$
and transverse flow velocity $\beta_T$ on collision energy
$\sqrt{s_{NN}}$, the excitation functions of $T_0$ and $\beta_T$
for central Au-Au collisions are shown in Figs. 2(a) and 2(b)
respectively. The results for $\pi^+$, $K^+$, $p$, $K^0_S$,
$\phi$, $\Lambda$, $\bar\Xi^+$ ($\Xi^+$, $\Xi$) and $\bar
\Omega^+$ ($\Omega^-+\Omega^+$, $\Omega$) are represented by
different symbols marked in the panels. One can see that $T_0$ in
Fig. 2(a) and $\beta_T$ in Fig. 2(b) increase quickly at lower
energies from 2.7 to 7.7 GeV due to the fact that the system got
higher excitation degree and stronger squeeze and expansion
degree. They remain constant from 7.7 to 39 GeV and then increase
up to higher energies. The variation of $T_0$ and $\beta_T$ at
different collision energies is displayed in Fig. 2(c). A larger
$T_0$ can be clearly seen at larger $\beta_T$ due to higher
collision energy, which shows positive correlation between $T_0$
and $\beta_T$.

We notice that there is a saturation in the excitation functions
of $T_0$ and $\beta_T$ in the BES energy range at the RHIC, which
means that the interaction mechanism or evolution process in
$\sqrt{s_{NN}}=7.7$--39 GeV is different from that in
$\sqrt{s_{NN}}<7.7$ GeV and in $\sqrt{s_{NN}}>39$ GeV. In our
opinion, the system is baryon-dominated in $\sqrt{s_{NN}}<7.7$
GeV, in which there is no phase transition from hadron matter to
QGP due to small energy deposition. The system is meson-dominated
in $\sqrt{s_{NN}}>39$ GeV, in which the phase transition had
happened in whole volume due to large energy deposition. The
system starts its phase transition in part volume at
$\sqrt{s_{NN}}=7.7$ GeV and undergoes from baryon-dominated to
meson-dominated due to phase transition in larger and larger
volume in $\sqrt{s_{NN}}=7.7$--39 GeV. The onset energy of part
phase transition is 7.7 GeV and that of whole phase transition is
39 GeV. The critical energy range is from 7.7 GeV to 39 GeV.

It is known that the chemical freeze-out temperature $T_{ch}$ and
baryon chemical potential $\mu_B$ from the thermal and statistical
model~\cite{79,80}, can be parameterized as
\begin{align}
T_{ch} = \frac{T_{\lim}}{1+\exp[2.60-\ln(\sqrt{s_{NN}})/0.45]}
\end{align}
and
\begin{align}
\mu_B = \frac{a}{1+0.288\sqrt{s_{NN}}},
\end{align}
where $T_{\lim}=0.1584$ GeV, $a=1.3075$ GeV, and $\sqrt{s_{NN}}$
is in the units of GeV~\cite{81}. We have the critical range of
$T_{ch}=0.138$--0.158 GeV and $\mu_B=0.406$--0.107 GeV while
$\sqrt{s_{NN}}=7.7$--39 GeV. These values show the ranges of
$T_{ch}$ and $\mu_B$ in the critical energy range.

We noticed that it is indeed possible to observe that while going
from low to high temperature, a hadron gas becomes more and more
meson dominated. As mentioned in ref.~\cite{81a}, with the help of
the thermal model calculations, it is possible to plot the
normalized entropy density for mesons and baryons, and then one
may observe that the dominance changes at a cross-over temperature
of $T_{ch}\approx0.140$ GeV corresponding to $\mu_B \approx0.406$
GeV and $\sqrt{s}\approx9.3$ GeV which are almost exactly the
values for $T_{ch}$ and $\mu_B$ the present work gets.

It is noteworthy that the kinetic freeze-out temperature for the
emission of multi-strange particles is observed considerably
higher than those for the emissions of other identified particles,
which reveals a picture of separate freeze-out processes for
other identified and multi-strange particles. Meanwhile, the
transverse flow velocity of multi-strange particles is lower than
that of other identified particles. The reason behind the high
kinetic freeze-out temperature (low transverse flow velocity) may
be that the multi-strange hadrons can be left behind in the system
evolution process due to their large mass. This possibility is a
reflection of hydrodynamic behavior~\cite{82}, in which massive
particles are leaved due to their small velocity.

With the increase of collision energy, both the kinetic freeze-out
temperature $T_0$ and transverse flow velocity $\beta_T$ increase
or keep invariant if phase transition had happened in part volume.
There is a positive correlation between $T_0$ and $\beta_T$ when
we study them over a wide energy range. This renders that the
system stays at high excitation state and undergoes large squeeze
and expansion due to large energy deposition at high energy. This
work does not support the negative correlation between $T_0$ and
$\beta_T$ when we increase the energy, though negative correlation
can be explained as long lifetime (then low excitation) and large
squeeze and expansion. Deservedly, for a given spectrum, $T_0$ and
$\beta_T$ is negative correlation, which is not the case for
varying energy.

The values of $q$ extracted from the spectra of $\pi^+$, $K^+$,
$p$, $K^0_S$, $\phi$, $\Lambda$, $\bar\Xi^+$ ($\Xi^+$, $\Xi$) and
$\bar \Omega^+$ ($\Omega^-+\Omega^+$, $\Omega$) in central AA
collisions at different energies do not show particular behavior,
but hardly energy dependent or slightly change with energy. As an
entropy index, $q$ characterizes the degree of equilibrium of the
system. Generally, an equilibrium state corresponds to $q$ to be
1. The values of $q$ obtained in this work are approximately close
to 1, which renders that the system in the considered energy range
stays approximately in an equilibrium state or in a few local
equilibrium states. This also renders that the blast-wave fit is
approximately useable in this work.

However, it should be noted that the entropy index $q$ is a very
very sensitive quantity. A large $q$ which is not close to 1
results in a wide distribution, and a small $q$ which is close to
1 results in a narrow distribution. It is the fact that $q=1.01$
is not close enough to 1. So, it does not imply that the Tsallis
blast-wave fit is close to its Boltzmann-Gibbs counterpart if we
use the same $T_0$ and $\beta_T$ in the case of $q=1.01$. To
reduce the difference between the Tsallis blast-wave fit and its
Boltzmann-Gibbs counterpart, we need $q=1.0001$ or the one which
is closer to 1.
\\

{\section{Summary and conclusions}}

The main observations and conclusions are summarized here.

(a) The transverse momentum (mass) spectra of $\pi^+$, $K^+$, $p$,
$K^0_S$, $\phi$, $\Lambda$, $\bar\Xi^+$ ($\Xi^+$, $\Xi$) and $\bar
\Omega^+$ ($\Omega^-+\Omega^+$, $\Omega$) produced in central AA
collisions at mid-$y$ or mid-$\eta$ over an energy range from 2.7
GeV to 2.76 TeV have been studied by the blast-wave fit with
Tsallis statistics. The kinetic freeze-out temperature $T_0$ and
transverse flow velocity $\beta_T$ are extracted from the fit to
transverse momentum (mass) spectra.

(b) The excitation functions of $T_0$ and $\beta_T$ show that both
$T_0$ and $\beta_T$ increase sharply with the increase of
collision energy from 2.7 to 7.7 GeV. Then they remains invariant
from 7.7 to 39 GeV. At above 39 GeV, they show the trend of
increase. The three energy ranges have identifiable boundaries and
render three different interaction mechanisms or evolution
processes.

(c) The system is baryon-dominated from 2.7 to 7.7 GeV, in which
there is no phase transition from hadron matter to QGP due to low
energy deposition. The system starts its phase transition in part
volume at 7.7 GeV and undergoes from baryon-dominated to
meson-dominated due to phase transition in larger and larger
volume in 7.7--39 GeV. The system is meson-dominated at above 39
GeV, in which the phase transition had happened in whole volume.

(d) The onset energy of part phase transition from hadron matter
to QGP is 7.7 GeV and that of whole phase transition is 39 GeV.
The multi-strange and other identified particles shows separate
freeze-out process due to the difference in temperature and flow
velocity. From the refined structure, the multiple freeze-out
scenarios are also observed due to the mass dependent temperature
and flow velocity.
\\
\\
\\
{\bf Data availability}

The data used to support the findings of this study are included
within the article and are cited at relevant places within the
text as references.
\\
\\
\\
{\bf Ethical approval}

The authors declare that they are in compliance with ethical
standards regarding the content of this paper.
\\
\\
\\
{\bf Disclosure}

The funding agencies have no role in the design of the study; in
the collection, analysis, or interpretation of the data; in the
writing of the manuscript; or in the decision to publish the
results.
\\
\\
\\
{\bf Conflict of interest}

The authors declare that there are no conflicts of interest
regarding the publication of this paper.
\\
\\
\\
{\bf Acknowledgments}

This work was supported by the National Natural Science Foundation
of China under Grant Nos. 11575103, 11947418, and 11505104, the
Chinese Government Scholarship (China Scholarship Council), the
Scientific and Technological Innovation Programs of Higher
Education Institutions in Shanxi (STIP) under Grant No. 201802017,
the Shanxi Provincial Natural Science Foundation under Grant No.
201901D111043, and the Fund for Shanxi ``1331 Project" Key
Subjects Construction.
\\
\\

\end{multicols}
\end{document}